# Harmonizing the Generation and Pre-publication Stewardship of FAIR bioimage data

*Nikki Bialy 0000-0001-9681-9632 (1), Frank Alber 0000-0003-1981-8390 (2), Brenda Andrews 0000-0001-6427-6493 (3), Michael Angelo (4), Brian Beliveau 0000-0003-1314-3118 (5), Lacramioara Bintu 0000-0001-5443-6633 (4), Alistair Boettiger 0000-0002-3554-5196 (4), Ulrike Boehm 0000-0001-7471-2244 (6), Claire M. Brown 0000-0003-1622-663X (7), Mahmoud Bukar Maina 0000-0002-7421-3813 (8), James J. Chambers 0000-0003-3883-8215 (9), Beth A. Cimini 0000-0001-9640-9318 (10), Kevin Eliceiri 0000-0001-8678-670X (1, 11), Rachel Errington (12), Orestis Faklaris 0000-0001-5965-5405 (13), Nathalie Gaudreault 0000-0002-9220-5366 (14), Ronald N. Germain 0000-0003-1495-9143 (15), Wojtek Goscinski 0000-0001-6587-1016 (16), David Grunwald 0000-0001-9067-804X (17), Michael Halter 0000-0002-1628-324X (18), Dorit Hanein 0000-0002-6072-4946 (19), John W. Hickey 0000-0001-9961-7673 (20), Judith Lacoste 0000-0002-8783-8599 (21), Alex Laude 0000-0002-3853-1187 (22), Emma Lundberg (4, 23), Jian Ma 0000-0002-4202-5834 (24), Leonel Malacrida 0000-0001-6253-9229 (25), Josh Moore 0000-0003-4028-811X (26), Glyn Nelson 0000-0002-1895-4772 (22), Elizabeth Kathleen Neumann 0000-0002-6078-3321 (27), Roland Nitschke 0000-0002-9397-8475 (28), Shuichi Onami 0000-0002-8255-1724 (29), Jaime A. Pimentel 0000-0001-8569-0466 (30), Anne L. Plant 0000-0002-8538-401X (18), Andrea J. Radtke 0000-0003-4379-8967 (15), Bikash Sabata (31), Denis Schapiro 0000-0002-9391-5722 (32), Johannes Schöneberg 0000-0001-7083-1828 (33), Jeffrey M. Spraggins 0000-0001-9198-5498 (34), Damir Sudar 0000-0002-2510-7272 (35), Wouter-Michiel Adrien Maria Vierdag 0000-0003-1666-5421 (36), Niels Volkmann 0000-0003-1328-6426 (19), Carolina Wählby 0000-0002-4139-7003 (37), Siyuan (Steven) Wang 0000-0001-6550-4064 (38), Ziv Yaniv 0000-0003-0315-7727 (15) and Caterina Strambio-De-Castillia 0000-0002-1069-1816 (17)*

---

1) Morgridge Institute for Research, Madison, USA. 2) University of California Los Angeles, USA. 3) University of Toronto, Toronto, Canada. 4) Stanford University, Palo Alto, USA. 5) University of Washington, Seattle, USA. 6) Carl Zeiss AG, Oberkochen, Germany. 7) McGill University, Montreal, Canada. 8) University of Sussex, Sussex, UK and Yobe State University, Nigeria. 9) University of Massachusetts, Amherst, USA. 10) Broad Institute of MIT and Harvard, Imaging Platform, Cambridge, USA. 11) University of Wisconsin-Madison, Madison, USA. 12) Cardiff University, Cardiff, UK. 13) Univ. Montpellier, CNRS, INSERM, Montpellier, France. 14) Allen Institute for Cell Science, Seattle, USA. 15) National Institute of Allergy and Infectious Diseases, National Institutes of Health, Bethesda, USA. 16) National Imaging Facility, Brisbane, Australia. 17) UMass Chan Medical School, Worcester, USA. 18) National Institute of Standards and Technology, Gaithersburg, USA. 19) University of California, Santa Barbara, USA. 20) Duke University, Durham, USA. 21) MIA Cellavie Inc., Montreal, Canada. 22) Newcastle University, Newcastle upon Tyne, UK. 23) SciLifeLab, KTH Royal Institute of Technology, Stockholm, Sweden. 24) Carnegie Mellon University, Pittsburgh, USA. 25) Institut Pasteur de Montevideo, & Universidad de la República, Montevideo, Uruguay. 26) German BioImaging - Gesellschaft für Mikroskopie und Bildanalyse e.V., Constance, Germany. 27) University of California, Davis, Davis, USA. 28) University of Freiburg, Freiburg, Germany. 29) RIKEN Center for Biosystems Dynamics Research, Kobe, Japan. 30) Universidad Nacional Autónoma de México, Cuernavaca, México. 31) Altos Labs, Redwood City, USA. 32) Heidelberg University Hospital, Heidelberg, Germany. 33) University of California San Diego, San Diego, USA. 34) Vanderbilt University School of Medicine, Nashville, USA. 35) Quantitative Imaging Systems LLC, Portland, USA. 36) European Molecular Biology Laboratorium, Heidelberg, Baden-Württemberg, Germany. 37) Uppsala University, Uppsala, Sweden. 38) Yale University, New Haven, USA

## Abstract

Biological imaging, combined with molecular insights into genes and proteins, holds immense promise for deepening our understanding of complex cellular systems and accelerating the development of predictive, personalized therapies for human health. To fully realize this potential at scale—and harness the power of AI/ML to extract novel biological insights and therapeutic interventions—it is necessary to transition from siloed datasets to globally shared, rigorously annotated, and computationally ready image data. This demands systematic harmonization of multidimensional bioimaging data, where interoperable formats, and standardized context-rich annotation, quality controls, and analytical pipelines transform scattered observations into a coherent knowledge base ripe for computational mining. Only this machine-actionable aggregation can provide the substrate for AI/ML to extract mechanistic insights into fundamental biological mechanisms, novel diagnostic biomarkers and intervention targets. Enabling seamless image data sharing in the life sciences requires addressing two key areas. The first is outlined in an accompanying publication, *Enabling Global Image Data Sharing in the Life Sciences*, which focuses on the publicly available repositories needed to share digital array data [1].

This White Paper details a comprehensive set of requirements for integrated image data and metadata management - from acquisition through dissemination - ensuring the contextual information necessary for assessing quality, interpreting scientific validity, and enabling meaningful reuse remains intrinsically linked to the data throughout its lifecycle. Critically, it recognizes that generating harmonized, well-annotated publicly available corpora of FAIR bioimage data requires these datasets to be "FAIR-from-the-start" - an objective that can only be achieved by enabling experimental scientists to manage, organize, and analyze their data according to community standards from the very first experiment.

Building on recent progress made by the bioimaging field towards establishing shared practices for bioimaging Quality Control (QC) and metadata capture, we present actionable recommendations to advance these efforts through embedding researcher-friendly integrated software infrastructure directly into pre-publication workflows, thus transforming disorganized data capture into structured, shareable resources ready for aggregation and reuse. Our ultimate goal is to expand the use of streamlined tools and practices thus transforming how researchers capture, annotate, analyze and eventually publish bioimaging data thus laying the foundation for a new era of data-driven discovery.

## Background and Motivation

Biological images provide unique temporal and spatial data about biological molecules, which significantly enhances our understanding of complex biological systems. Moreover, sharing the vast amount of information captured in biological image data promises to advance human health and the economy, the same way sharing of genomics data and protein structure data has (i.e., the Human Genome Project, https://www.genome.gov/human-genome-project; and the Protein Data Bank - PDB, https://www.rcsb.org). Achieving this vision will require the interoperability, integration, and sharing of bioimage data across laboratories and research studies, thus maximizing the value of the billions of dollars invested annually in research around the globe. Data sharing is globally recognized as highly desirable [2,3], but too often, preclinical bioimaging data remains inaccessible, because standardized practices and consistent metadata for acquisition, analysis and publication are not yet employed/enforced. Moreover, even when shared, bioimaging data often lacks sufficient information to evaluate its validity and reproducibility, to assess its reusability, drive machine-actionable aggregation and ultimately provide the substrate for AI/ML to extract scientific and therapeutic advances. [4–12].

Funding agencies have been working to improve the situation, for example, by establishing data governance and preservation policies [13–15] that make it mandatory for researchers to provide detailed data management and sharing plans [13,16–18]. To improve reproducibility in research, many publishers now require more detailed materials and methods sections [6,19–22]. In 2016, stakeholders from academia, industry, funding agencies, and publishing arrived at a set of FAIR Data Principles—"Findable, Accessible, Interoperable and Reusable"—which are expected to improve the machine-actionability of data thus expanding its utility well beyond its original purpose [23]. Several international initiatives are engaged in addressing the challenges associated with the implementation of FAIR principles during bioimage data generation and post-acquisition processing (Supplemental Table 1). For many years, the Open Microscopy Environment (OME) has produced tools, metadata models and shared bioimage data file formats to facilitate the generation, pre-publication management, integrated analysis and post-publication sharing of contextual-information rich image data [5,24–29]. Organizations of bioimaging experts across academia and industry (see Supplemental Table 1) have more recently coalesced around improving standard practices for instrument QC and metadata [8,20,30–43] Through these multi-stakeholder collaborations, individual laboratories and core facilities are partnering with manufacturers to develop shared metadata standards and conduct inter-laboratory studies validating quality control methods [44–50]. Despite limited funding, these groups have driven progress through voluntary commitments, enabling tangible—if still limited—progress across the bioimaging field. While challenges remain, these efforts lay the foundation for generating reliable, well-documented bioimaging data that meets FAIR principles and are critical to unlocking the full potential of quantitative bioimage-based research and its biomedical applications.

While most scientists endorse FAIR principles and intend to share bioimage data, persistent technical barriers hinder implementation. The first challenge involves the lack of federated public bioimage data repositories spanning individual institutions to the national level [1,51–53]. The second stems from fragmented practices in metadata annotation, bioimage acquisition, analysis, and dissemination - where lack of locally-available and usable software platforms, inconsistent standards and poor machine-readability create bottlenecks in automation and prevent the generation of bioimage datasets that are ready to be shared. This fragmentation particularly impacts pre-publication data management, where the absence of integrated software platforms and unified workflows available to researchers at the local level, limits both immediate analysis and long-term reuse potential. Addressing these gaps requires user-friendly tools that not only integrate with experimental workflows, [54–56] but also bridge the divide between local data production and analysis, and broader research cyberinfrastructure.

Here we summarize recent progress toward standardized data and metadata practices for bioimages, while recommending how to extend this work to address remaining challenges. Our goal is to make software platforms, best practices and usable tools accessible to all biomedical researchers, regardless of their expertise, available resources, or geographic location. We adopt a comprehensive framework that follows the bioimage-data lifecycle (Figure 1). Beginning with data generation, where challenges include sample preparation and bioimage acquisition, we then examine issues affecting the reliability and reproducibility of post-acquisition processing, visualization, and quantitative analysis. Finally, we address the critical importance of data management and stewardship in maintaining robust links between bioimage data and metadata throughout the entire research process and often culminating in publication. Building integrated software platforms that ensure complete provenance tracking and quality assessment, forms the essential foundation for achieving truly FAIR-compliant microscopy data. Only through such comprehensive approaches can we maximize the scientific value of bioimage data, both for its original research purpose and for enabling new discoveries through data sharing and reuse.

## Challenges and Recent Progress Associated with Data Generation

The preservation of data and metadata starting from data generation lies at the very heart of the bioimage data lifecycle, involves considerations made at the planning phase of the research project, is relevant before the sample hits the bioimage acquisition platform, informs all subsequent bioimage processing steps and is essential to ensuring the scientific value of shared bioimage data.

Good practices in data generation and management that are deployed throughout the entire data life-cycle and ensure that data are "FAIR from the start" are essential for rigorous and reproducible quantitative bioimage-based biomedical research and for producing bioimage data that can be interpreted, trusted, and reused through model-based and data-driven mining, aggregation, reanalysis, and integrative modeling. It is crucial that third-party data users have ready access not only to the bioimages themselves, but crucially to machine-actionable representations of all data-related information (i.e., "bioimage metadata") that allows to evaluate the suitability of given datasets for answering specific scientific questions before accessing or downloading them. This "bioimage metadata" comprises two types of metadata[31]: (i) *data provenance metadata*, which describes the experimental conditions, sample description and preparation, bioimage acquisition (i.e., hardware description and bioimage acquisition settings), and bioimage processing, visualization, and analysis; and (ii) *QC metadata*, which describes system performance recorded through standardized QC protocols and metrics.

### <u>Advances in Capturing Experimental Conditions and Sample Preparation Metadata</u>

Accurate interpretation of microscopy data requires comprehensive documentation of the entire experimental workflow. This includes: (1) all pre-imaging procedures, (2) biological sample origin and preparation methods, (3) labelling protocols and reagents, (4) mounting techniques, and (5) characteristics of specimen holders during bioimaging. While traditionally reported in Methods sections [4,19,20,50,57], this information must now be captured and retained with the bioimage, as structured, machine-readable metadata to enable data reuse. The bioimaging field is making significant strides toward standardization. The REMBI framework[33], developed through multi-stakeholder consensus, establishes core metadata categories for light, electron, and X-ray microscopy to ensure data interpretability and reusability. Along the same lines, minimum information guidelines for 3D microscopy, highly multiplexed tissue images and cell migration experiments have recently been developed to guide the documentation of experimental procedures and sample preparation [58–62] Effective implementation requires three key components: standardized ontologies for consistent knowledge representation [63–67], automated metadata capture tools [32,62,68–73], and seamless integration with data management platforms. Together, these elements create the foundation for FAIR-compliant microscopy research[25,74].

### **Advances in Capturing Microscopy Acquisition Provenance and QC Metadata**

Quality assessment, reproducibility, interpretation, and reuse of bioimage data depend on detailed information about hardware specifications, bioimage acquisition settings, and instrument performance at the time of data acquisition [30,31]. A comprehensive technical description of the bioimaging system configuration enables calculation of key parameters such as spatiotemporal resolution, system noise, and the physical and temporal dimensions of pixel data. An instrument performance assessment plan, including regular tracking of standardized QC metrics, quantifies performance changes over time [44–47]. These QC metrics compare expected (theoretical) values with observed (empirical) measurements and with baseline values recorded at instrument installation (time zero). Importantly, QC metrics also support characterization and calibration of derived bioimage analysis outputs, such as co-registration measurements. Capturing the overall state [32] and performance of the microscope as metadata at the time of acquisition is essential to identify potential batch effects in large datasets [9]. Such batch effects can significantly impact the performance of artificial intelligence and machine learning algorithms, making it critical to document instrument state and performance for accurate result interpretation [9,10].

## Ensuring the Validity and Reproducibility of Image Data Visualization, and Analysis

A detailed workflow must be provided to independently assess the quality, reproducibility, and interpretation of bioimage data analysis, as outlined in recent community guidelines [19,75,76]. Analysis metadata should also include data size, computing hardware specifications, and networking requirements. This is essential, as data analysis pipelines typically involve complex dependency chains of multiple software packages (e.g., open source software tools such as FiJi, QUPath and napari, etc.; commercial tools such as Imaris, Huigyens, etc.) that must be precisely described to ensure reproducibility. Consequently, the metadata associated with the analysis phase, must specify the exact versions of all software components to prevent discrepancies caused by subsequent software updates. Manually reporting all dependencies beyond the primary software is impractical and is best managed using automated package managers (e.g., the pip freeze command).

Since image processing pipelines often involve multiple tools, it is essential to store intermediate and final results, along with associated metadata, in a harmonized and comparable format across different software tools [77]. To address this, developers have made significant progress in using containerization technologies [78–81] and workflow management tools [82–87]. Future efforts should focus on enhancing the robustness of these solutions and encouraging their widespread dissemination and adoption.

If an analysis environment is recreated using the same hardware, operating system, bioimage analysis software, and parameter settings, analyzing the same dataset at different locations should yield the same results. However, in the case of algorithms that rely on randomness, results within an acceptable margin of error are considered sufficiently similar [88–90]. In addition such cases require sharing additional details such as random seed values and other relevant software parameters. For instance, in deep learning, reproducing results from a retrained or new model necessitates access to both the code and model weights, which should be shared in interoperable formats like the Open Neural Network Exchange (ONNX).

## Everyday Stewardship of Bioimage Data and Metadata During Active Data Production

Data stewardship is an intrinsic and essential aspect of generating high-quality bioimage data that is "FAIR from the start." For this to happen, data stewardship must involve the entire lifecycle of the data, starting with the planning phase and continuing during experimental design and execution, sample preparation, data acquisition, post-processing, visualization, and analysis. In addition, data stewardship must outlive the research project to ensure that well-documented published data remains available for re-use, as detailed in the companion manuscript [91]. Specifically, correct data stewardship ensures that the conditions used to generate, process, analyze, and validate data are transparent, documented, propagated alongside the data and automatically reported to downstream users in both human-readable forms (i.e., scientific publications) [4,20,21,57] and structured machine-readable metadata frameworks [92–94]. Cloud-ready data exchange formats [28,29,42] and standardized Application Programming Interfaces (APIs) are necessary for integration of bioimages and results [28–30,32,33,58,95]. This, in turn, ensures that data can be trusted, correctly interpreted, reproduced, and reused through data aggregation, mining, integrative modeling, and further analysis (including AI/ML). Proper data stewardship is crucial from the start, to avoid wasting time and resources later trying to re-generate data that has been lost or cannot be interpreted, and consequently, promote efficiency and sustainability (economic, environmental, and societal) [96,97].

As such, effective data stewardship requires well-maintained, enterprise-grade, scalable, open-source, and democratized Research Data Management and Sharing (RDMS) cyberinfrastructure, capable of providing the necessary pipelines to move data along from generation to analysis and sharing [51] (Figure 1). This cyberinfrastructure should include the following; 1) Persistent Identifiers (PIDs) for research resources, individuals, publications and data [98–101]; 2) shared file formats and associated APIs [28,29,102]; 3) the use of community-defined ontologies [103,104105,106]; 3) bioimage metadata specifications [30,33,58]; and 4) community defined Next Generation Metadata frameworks [95]. This cyberinfrastructure should interface with Electronic Lab Notebooks (ELN) and Laboratory Information Management Systems (LIMS) and, whenever possible, automatically capture and propagate output metadata from all relevant instrumentation (including but not limited to robotic, microfluidics, and bioimage acquisition hardware) [107,108]. In summary, cyberinfrastructure should cover the following three interconnected aspects:

- **WHAT information should be captured in Bioimage Metadata:** develop community standards models for all aspects of image metadata [30,31,33,58].
- **WHERE Bioimage Metadata should be stored:** develop community specifications and shared APIs (i.e., OME-NGFF and Next Generation Metadata)[28,29,95,109] to ensure data and metadata are stored in synch with each other across the bioimaging data flow.
- **HOW Bioimage Metadata should be captured:** develop easy to use tools to facilitate metadata annotation, data curation, and seamless integration of all aspects of the bioimaging pipeline (i.e., integration with LIMS, ELNs, and hardware instrumentation; ontology-enriched, REMBI-based, modular template spreadsheets; ad hoc graphical interfaces to collect microscopy metadata) [32,69,71,72,110–114].

## FAIR quality control in Recently Emerging Bioimaging Modalities

A set of recently developed bioimaging modalities are emerging as techniques of choice to quantify the spatial distribution of molecules and supramolecular structures at the subcellular, cellular, tissue, and organismal levels as reviewed in [115–117118–120]. These techniques include spatial RNA profiling methods capable of resolving hundreds of probes at subcellular resolution using light microscopy, *in situ* sequencing (ISS), and capture assay. In addition, several highly multiplexed antibody-based bioimaging techniques have also been developed and commercialized. Cryogenic transmission electron microscopy (cryo-EM) provides images of biological matter in a frozen-hydrated, near-native state. The cryo-EM field is recognized as a good example of effectively annotating metadata and stewardship of raw and derived data [33]. The cryo-EM community agrees that detailed metadata must be made publicly available alongside the data and that metadata standards must be reviewed regularly to ensure fitness and relevance to the evolving community needs [33,121]. Data sharing of cryo-EM derived [122–124]) and raw data [125,126] - both with associated metadata - has already been established and is enforced by most publishers,

We advocate for a similar approach to be adopted for multiplexed RNA, protein, and multi-omic bioimaging datasets. The recent introduction of these methods, combined with their complexity and diversity, resulted in various workflows related to sample processing, reagent QC, bioimage acquisition, bioimage processing, and bioimage analysis without agreed-upon community standards [127]. The establishment and wide adoption of such standards are essential for cross-lab and cross-platform data interoperability and analysis, which is even more critical to the community since these data are expensive to acquire [116,118,128] and are associated with a number of consortia efforts [129–131]. Critical details related to sample preparation, reagent validation, and platform-specific bioimaging parameters must be recorded and shared using community-defined metadata. We highlight specific challenges around harmonizing multiplexed image processing and reporting that limit current ability towards FAIR data [23].

## Recommendations Toward Global Bioimage Data Generation and Stewardship Standards

All stakeholders have a role to play in implementing these practices. Funding agencies should support the development of tools, protocols, and metadata standards by bioimaging communities to ensure their routine adoption by core facilities and users. Manufacturers can facilitate automated access to comprehensive, community-defined technical descriptions of instruments and quality control (QC) metrics within the metadata they provide with the bioimage data [107,108]. Publishers ought to require that microscopy metadata—including hardware specifications, bioimage acquisition settings, and QC metrics—be included as part of data submissions. Investigators should embrace best practices established by the bioimaging research community, providing both human- and machine-readable metadata to ensure their data meet FAIR principles. Finally, fostering the creation, dissemination and support of resources in multiple languages is essential for universal adoption of these standards.

In Text Box 1, we outline recommended steps to promote the generation and stewardship of image data that is “FAIR from the start,” ready for sharing and reuse. Additional details are available in Supplemental Material.

## Figures

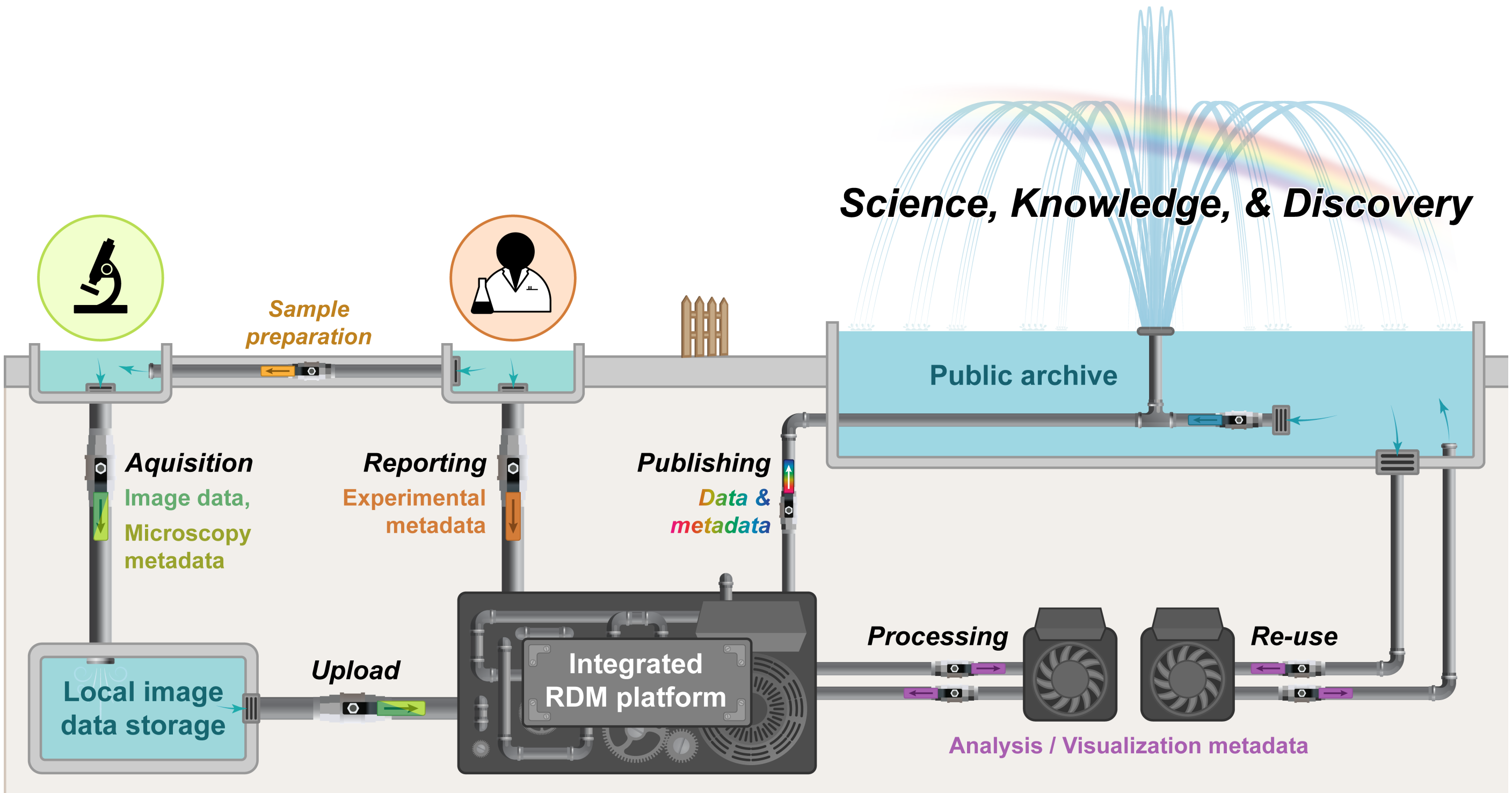

**Figure 1: Research Data Management and Sharing for images: beautiful fountains require ugly piping!** Sharing research data enhances transparency, collaboration, efficiency, and impact. Image data, however, poses distinct challenges: it comes from diverse sources and rapidly evolving technologies that produce massive, multimodal datasets. While the value of scientific discoveries and medical advances (top right, "beautiful fountain") is easily appreciated, the critical “plumbing” (bottom) of RDMS infrastructure that makes them possible often goes overlooked. The figure illustrates the *key interrelated steps* (represented as reservoirs and engines) required to produce high-quality bioimage data that is ready to share in compliance Findable, Accessible, Interoperable and Reusable (FAIR) principles. These steps include experimental setup, sample preparation, and image acquisition, analysis and visualization and are often carried out by different researchers requiring the integrated management of *data and metadata* (represented as water flowing through pipes connecting the different steps) to ensure image data can be trusted, results properly interpreted and *scientific advancements* (represented as beautiful water works) achieved. Figure credit: Matthew et al., 2024 (https://doi.org/10.5281/zenodo.14020675).

**Text Box 1 - A to-do list for various stakeholders**

## Towards Global Image Data Generation and Stewardship

## A to-do list for various stakeholders

More details on the action items below are provided in Supplemental Material.

1. Ensure the **long-term sustainability of national and international bioimaging communities** (e.g., ABIC, ABRF, BINA, CBI, GBI, LABI, and QUAREP-LiMi; see also Table 1 in supplemental materials), with the specific purpose of enabling recurring gatherings to coordinate (i.e., discuss, recommend, update) the development of:
   a. **Consensus guidelines for quality control procedures and standards** to encourage the implementation and reporting of QC protocols and performance benchmarks, including for imaging instrumentation.
   b. **Shared metadata specifications, exchange frameworks, and tools** to minimize barriers to adopting metadata guidelines for academic, government, and industry stakeholders.
   c. **Shared computational cyberinfrastructure is needed to generate and manage image data before publication.** This infrastructure should include well-documented and high-speed software tools, frameworks, computing and storage equipment, and networks. It will support all stages of the imaging process, from data annotation to image acquisition and analysis. Community-defined standards are essential to ensure transparency about the instruments and algorithms used.
2. **Invest in core facilities (aka Shared Research Resources) and their Personnel from all backgrounds and regions** to:
   a. **Provide expertise** in sample preparation, validation of staining protocols, image acquisition, and image analysis.
   b. **Democratize access** through shared resources and promote collaborations to facilitate access to advanced technology.
   c. Serve as **pivotal hubs for disseminating expertise and user training** on all topics essential for preparing FAIR image data for sharing and engendering maximum reuse value across resourced and under-resourced regions and communities.
   d. **Develop strong connections with software development centers** to ensure cyberinfrastructure's usability, customization, and democratization for imaging pipeline automation.
3. Support the **career and recognition of imaging scientists** specializing in generating and stewardship of FAIR image data. These include core facility personnel, image data curators, bio-image analysis experts, and research software engineers.
4. In collaboration with vendors, **develop and deploy automated methods to capture harmonized and consistent metadata documenting all steps of the imaging pipeline,** from reagents used to generate image data to microscopy instruments and peripherals.
5. **Promote the use of Persistent Identifiers (PID) for the FAIR description of research resources** (e.g., reagents, instruments, core facilities) and outputs (e.g., datasets) to facilitate reproducibility and reuse and ensure that the personnel involved in the research enterprise are appropriately acknowledged.
6. **Develop metrics that describe the qualities of resultant image data**.

## List of Contributors

| Name | Email | ORCID | Affiliation | Funding Statement | Contribution* |
|---|---|---|---|---|---|
| Nikki Bialy | nbialy@morgridge.org | https://orcid.org/0000-0001-9681-9632 | BioImaging North America (BINA RRID: SCR_024409), Morgridge Institute for Research, 330 North Orchard Street, Madison, WI 53715, USA. | NB is supported through a grant from the Chan Zuckerberg Initiative DAF, an advised fund of Silicon Valley Community Foundation, to the Morgridge Institute for Research for BioImaging North America (BINA). | Resources, Writing - Original Draft, Writing - Review & Editing, Endorsement, Supervision, Project Administration |
| Frank Alber | falber@g.ucla.edu | https://orcid.org/0000-0003-1981-8390 | Department of Microbiology, Immunology & Molecular Genetics, University of California Los Angeles, USA | | Endorsement |
| Brenda Andrews | brenda.andrews@utoronto.ca | https://orcid.org/0000-0001-6427-6493 | The Donnelly Centre, University of Toronto, Toronto Canada M5S 3E1 | Work in the Andrews lab is supported by the NIH (R01HG005853) and the Canadian Institutes of Health Research (CIHR). BA holds at Tier 1 Canada Research Chair in Systems Genetics and Cell Biology | Writing - Original Draft, Endorsement |
| Michael Angelo | mangelo0@stanford.edu | | Stanford University School of Medicine, Palo Alto, CA, USA | | Writing - Original Draft |
| Brian Beliveau | beliveau@uw.edu | https://orcid.org/0000-0003-1314-3118 | University of Washington | BJB acknowledges support from the National Institutes of Health under grant 1R35GM137916 (NIGMS). | Writing - Original Draft |
| Lacramioara (Lacra) Bintu | lbintu@stanford.edu | https://orcid.org/0000-0001-5443-6633 | Stanford University, Stanford, CA, USA | | Endorsement |
| Alistair Boettiger | aboettig@stanford.edu | https://orcid.org/0000-0002-3554-5196 | Stanford University, Stanford, CA, USA | | Writing - Review & Editing, Endorsement |
| Ulrike Boehm | ulrike.boehm@gmail.com | https://orcid.org/0000-0001-7471-2244 | Carl Zeiss AG, Oberkochen, Germany | | Writing - Review & Editing, Endorsement |
| Claire M. Brown | claire.brown@mcgill.ca | https://orcid.org/0000-0003-1622-663X | Advanced BioImaging Facility (ABIF), McGill University, Montreal, Quebec, H3G 0B1, Canada | This project has been made possible in part by grant number 2020-225398 from the Chan Zuckerberg Initiative DAF, an advised fund of Silicon Valley Community Foundation. | Endorsement |
| Mahmoud Bukar Maina | M.Bukar-Maina@sussex.ac.uk | https://orcid.org/0000-0002-7421-3813 | University of Sussex, Sussex, UK & Biomedical Science Research and Training Centre, Yobe State University, Nigeria | | Writing - Review & Editing, Endorsement |
| James J. Chambers | jjchambe@umass.edu | https://orcid.org/0000-0003-3883-8215 | Institute for Applied Life Sciences, University of Massachusetts, Amherst, MA 01003, USA | | Endorsement |
| Beth A. Cimini | bcimini@broadinstitute.org | https://orcid.org/0000-0001-9640-9318 | Broad Institute of MIT and Harvard, Imaging Platform, Cambridge, MA, USA | This publication has been made possible in part by CZI grant 2020-225720 (DOI:10.37921/977328pjvbca) from the Chan Zuckerberg Initiative DAF, an advised fund of Silicon Valley Community Foundation (funder DOI 10.13039/100014989. This work was also supported by the Center for Open | Writing - Review & Editing, Endorsement |
| | | | | Bioimage Analysis (COBA) funded by the National Institute of General Medical Sciences P41 GM135019 | |
| Kevin Eliceiri | eliceiri@wisc.edu | https://orcid.org/0000-0001-8678-670X | Morgridge Institute for Research and the University of Wisconsin-Madison | This work was supported by the Center for Open Bioimage Analysis (COBA) funded by the National Institute of General Medical Sciences P41 GM135019 (Cimini and Eliceiri) | Writing - Review & Editing, Endorsement |
| Rachel Errington | ErringtonRJ@cardiff.ac.uk | | School of Medicine, Cardiff University, Cardiff, UK | | Writing - Original Draft, Writing - Review & Editing |
| Orestis Faklaris | orestis.faklaris@mri.cnrs.fr | https://orcid.org/0000-0001-5965-5405 | BCM, Univ. Montpellier, CNRS, INSERM, Montpellier 34293, France | OF is supported by the French National Research Agency (ANR-10-INBS-04). | Writing - Review & Editing, Endorsement |
| Nathalie Gaudreault | nathalieg@alleninstitute.org | https://orcid.org/0000-0002-9220-5366 | Allen Institute for Cell Science, Seattle, WA, USA | We wish to thank the Allen Institute for Cell Science Founder, Paul G. Allen, for his vision, encouragement, and support. | Writing - Review & Editing, Endorsement |
| Ronald N. Germain | rgermain@niaid.nih.gov | https://orcid.org/0000-0003-1495-9143 | Laboratory of Immune System Biology, NIAID, NIH | Intramural Program of NIAID, NIH. | Writing - Review & Editing, Endorsement |
| Wojtek Goscinski | w.goscinski@anif.org.au | https://orcid.org/0000-0001-6587-1016 | National Imaging Facility, Brisbane, Australia | The authors acknowledge the facilities and scientific and technical assistance of the National Imaging Facility, a National Collaborative | Endorsement |
| | | | | Research Infrastructure Strategy (NCRIS) capability | |
| David Grunwald | David.Grunwald@umassmed.edu | https://orcid.org/0000-0001-9067-804X | RNA Therapeutics Institute, UMass Chan Medical School, Worcester MA 01605, USA | | Writing - Review & Editing, Endorsement |
| Michael Halter | michael.halter@nist.gov | https://orcid.org/0000-0002-1628-324X | National Institute of Standards and Technology, Gaithersburg, MD, USA | | Writing - Review & Editing, Endorsement |
| Dorit Hanein | dorit@ucsb.edu | https://orcid.org/0000-0002-6072-4946 | Departments of Biochemistry and Chemistry, of Biological Engineering, University of California, Santa Barbara, CA, USA | DH research is in part sponsored by the U.S. Army Research Office and accomplished under contract W911NF-19-D-0001 for the Institute for Collaborative Biotechnologies. | Writing - Original Draft, Writing - Review & Editing, Endorsement |
| John W. Hickey | john.hickey@duke.edu | https://orcid.org/0000-0001-9961-7673 | Department of Biomedical Engineering, Duke University Durham, North Carolina, USA | | Writing - Original Draft, Endorsement |
| Judith Lacoste | jlacoste@miacellavie.com | https://orcid.org/0000-0002-8783-8599 | MIA Cellavie Inc., Montreal, Quebec, H1K 4G6, Canada | | Endorsement |
| Alex Laude | alex.laude@newcastle.ac.uk | https://orcid.org/0000-0002-3853-1187 | BioImaging Unit, Newcastle University, UK | Newcastle University | Endorsement |
| Emma Lundberg | emmalu@stanford.edu | | Stanford University, California, USA and SciLifeLab, KTH Royal Institute of | | Endorsement |
| | | | Technology, Stockholm, Sweden | | |
| Jian Ma | jianma@cs.cmu.edu | https://orcid.org/0000-0002-4202-5834 | Carnegie Mellon University, Pittsburgh, PA, USA | | Endorsement |
| Leonel Malacrida | lmalacrida@pasteur.edu.uy | https://orcid.org/0000-0001-6253-9229 | Institut Pasteur de Montevideo, & Universidad de la República, Montevideo, Uruguay | | Endorsement |
| Josh Moore | josh@openmicroscopy.org | https://orcid.org/0000-0003-4028-811X | German BioImaging-Gesellschaft für Mikroskopie und Bildanalyse e.V., Constance, Germany | JM is supported by the Deutsche Forschungsgemeinschaft (DFG, German Research Foundation) – 501864659 as part of NFDI4BIOIMAGE. | Writing - Review & Editing, Endorsement |
| Glyn Nelson | glyn.nelson@ncl.ac.uk | http://orcid.org/0000-0002-1895-4772 | Bioimaging Unit, Newcastle University, Newcastle upon Tyne, NE2 4HH, UK | Newcastle University, UK | Writing - Review & Editing, Endorsement |
| Elizabeth Kathleen Neumann | ekneumann@ucdavis.edu | https://orcid.org/0000-0002-6078-3321 | Department of Chemistry, University of California, Davis, Davis, California, USA | University of California, Davis | Endorsement |
| Roland Nitschke | Roland.Nitschke@biologie.uni-freiburg.de | https://orcid.org/0000-0002-9397-8475 | Life Imaging Center, Signalling Research Centres CIBSS and BIOSS, University of Freiburg, 79104 Freiburg, Germany | R.N. was supported by grant NI 451/10-1 from the German Research Foundation and grant 03TN0047B 'FluMiKal' from the German Federal Ministry for Economic Affairs and Climate Action | Writing - Review & Editing, Endorsement |
| Shuichi Onami | sonami@riken.jp | https://orcid.org/0000-0002-8255-1724 | RIKEN Center for Biosystems Dynamics Research, Kobe, Japan | | Endorsement |
| Jaime A. Pimentel | arturo.pimentel@ibt.unam.mx | https://orcid.org/0000-0001-8569-0466 | Laboratorio Nacional de Microscopía Avanzada, Instituto de Biotecnología, Universidad Nacional Autónoma de México, Cuernavaca, Morelos, 62210, México | | Writing - Review & Editing, Endorsement |
| Anne L. Plant | anne.plant@nist.gov | https://orcid.org/0000-0002-8538-401X | National Institute of Standards and Technology, Gaithersburg, MD, USA | | Writing - Review & Editing, Endorsement |
| Andrea J. Radtke | andrea.radtke@nih.gov | https://orcid.org/0000-0003-4379-8967 | Laboratory of Immune System Biology, Lymphocyte Biology Section and Center for Advanced Tissue Imaging, NIAID, NIH, Bethesda, MD, USA | AJR is supported by the Intramural Research Program of the NIH, National Institute of Allergy and Infectious Diseases (NIAID) and National Cancer Institute (NCI) | Writing - Original Draft, Writing - Review & Editing, Endorsement |
| Bikash Sabata | bsabata@altoslabs.com | NA | Altos Labs, Redwood City, CA, USA | NA | Writing - Review & Editing, Endorsement |
| Denis Schapiro | Denis.Schapiro@uni-heidelberg.de | https://orcid.org/0000-0002-9391-5722 | Institute for Computational Biomedicine, Heidelberg University Hospital, Heidelberg, Germany; Institute of Pathology, Heidelberg University Hospital, Heidelberg, Germany; Translational | DS is supported by the German Federal Ministry of Education and Research (BMBF 01ZZ2004); the Ministry for Science, Research and Science Baden-Württemberg „AI Health Innovation Cluster“ and "MULTI-SPACE"; Bruno und Helene Joester Stiftung and research funding from Cellzome, a GSK company. | Writing - Original Draft, Writing - Review & Editing, Endorsement |
| | | | Spatial Profiling Center (TSPC), Heidelberg, Germany | | |
| Johannes Schöneberg | jschoeneberg@health.ucsd.edu | https://orcid.org/0000-0001-7083-1828 | University of California, San Diego, CA, USA | JS is supported by the Hartwell Foundation, the W.M.Keck Foundation, the Brain Research Foundation, The Chan Zuckerberg Foundation, and the NIH (1DP2GM150022-01, 1R01GM148765-01). | Writing - Review & Editing, Endorsement |
| Jeffrey M. Spraggins | jeff.spraggins@Vanderbilt.Edu | https://orcid.org/0000-0001-9198-5498 | Department of Cell & Developmental Biology, Vanderbilt University School of Medicine, Nashville, TN, USA | J.M.S. is supported by the National Institutes of Health U01DK133766 (NIDDK), U54DK134302 (NIDDK), U54EY032442 (NEI), R01AG078803 (NIA), R01AI145992 (NIAID), and R01AI138581 (NIAID). | Writing - Review & Editing |
| Damir Sudar | dsudar@qitissue.com | https://orcid.org/0000-0002-2510-7272 | Quantitative Imaging Systems LLC, Portland, OR, USA | DS is supported by the National Institutes of Health U2CCA23380 (NCI) and 1 R44 CA250861 (NCI). | Endorsement |
| Wouter-Michiel Adrien Maria Vierdag | michiel.vierdag@embl.de | https://orcid.org/0000-0003-1666-5421 | Genome Biology Unit, European Molecular Biology Laboratorium, Heidelberg, Baden-Württemberg, Germany | WMV receives research funding from Cellzome, a GSK company. | Writing - Original Draft, Writing - Review & Editing |
| Niels Volkmann | nvo@ucsb.edu | https://orcid.org/0000-0003-1328-6426 | Departments of Bioengineering, of Biological Engineering, Electrical and Computer Engineering, and Biomolecular Science and Engineering Program, University of California, Santa Barbara, CA, USA | | Writing - Original Draft, Writing - Review & Editing, Endorsement |
| Carolina Wählby | carolina.wahlby@it.uu.se | https://orcid.org/0000-0002-4139-7003 | Dept. Information Technology and Science for LifeLaboratory, Uppsala University, Uppsala, Sweden | | Writing - Review & Editing, Endorsement |
| Siyuan (Steven) Wang | siyuan.wang@yale.edu | https://orcid.org/0000-0001-6550-4064 | Yale University | | Endorsement |
| Ziv Yaniv | zivyaniv@nih.gov | https://orcid.org/0000-0003-0315-7727 | Bioinformatics and Computational Bioscience Branch, National Institute of Allergy and Infectious Diseases, National Institutes of Health, Bethesda, MD, USA | Z.Y. is supported by the Bioinformatics and Computational Biosciences Branch (BCBB) Support Services Contract HHSN316201300006W/75N93022F00001 to Guidehouse Inc. | Writing - Original Draft, Writing - Review & Editing, Endorsement |
| Caterina Strambio-De-Castillia | caterina.strambio@umassmed.edu | https://orcid.org/0000-0002-1069-1816 | Program in Molecular Medicine, UMass Chan Medical School, Worcester MA | | Conceptualization, Methodology Resources, Writing - Original Draft, Writing - Review & Editing, Endorsement, |
| | | | 01605, USA | | Supervision, Funding acquisition |

* Author contributions categories comply with the CRediT initiative [132]

## Acknowledgements

Disclaimer: Commercial products are identified in this document in order to specify the experimental procedure adequately. Such identification is not intended to imply recommendation or endorsement by the National Institute of Standards and Technology, nor is it intended to imply that the products identified are necessarily the best available for the purpose. The Authors thank Darryl Conte Jr., Ph.D., for insightful comments and suggested revisions to the manuscript.

## Conflict of Interest Statements

Below are the statements shared by contributors indicating potential conflict of interest.

| Name | Statement |
|---|---|
| Ulrike Boehm | UB's contribution to this manuscript is a result of her voluntary involvement with QUAREP-LiMi and BINA, and does not reflect the position of Carl Zeiss AG on this matter. |
| Josh Moore | holds equity in Glencoe Software. |
| Denis Schapiro | DS reports funding from GSK and received honorariums from Immunai, Noetik, Alpenglow and Lunaphore. |
| Damir Sudar | DSu is employed by Quantitative Imaging Systems, a commercial entity developing imaging software. |
| Siyuan (Steven) Wang | Founder, shareholder, consultant of Translura, Inc |

## Supplemental Material

### Data Generation Recommendations

Overcoming challenges related to the generation of bioimage data that is "FAIR from the start" requires specific solutions that should be planned for and carried out by all interested stakeholders. To guide the development of these solutions, we provide the following specific recommendations:

- **Promote the widespread adoption of Persistent IDentifiers (PIDs)** to enable the FAIR description of all scientific entities, to facilitate reporting and reproducibility and to recognize the essential roles of bioimaging scientists and core facilities in the research enterprise [98–101]. This recommendation is backed by recent cost-benefit analyses [99,100] that revealed significant financial advantages associated with the adoption of PIDs which would be related to reductions in staff salaries, the time saved from tedious data entry, and in the facilitation of technology advances. Example of commonly used PIDs include:
    - Research Organization Registry (ROR) [133,134] for institutions and free standing core facilities.
    - Research Resource Identifier (RRID) [134,135] for rsearch core facilities that are part of larger institutions, reagents, research resources and instruments.
    - Digital Object Identifier (DOI) for datasets, publications and other digital resources.
    - Open Researcher and Contributor ID (ORCID) [136,137] for individuals.
    - PIDINST for instrument instances [101,138,139].

- **Promote the collection of full technical descriptions of microscope hardware specifications, image acquisition settings, and QC protocols and metrics** (aka Microscopy Metadata) that comply 4DN-BINA-OME (NBO-Q) microscopy metadata specifications [30,31] which is being developed by consensus by bioimaging scientists and instrument manufacturers [107,108]. Technical descriptions captured in Microscopy Metadata must become obligatory aspects of the production of any image data; in their absence, image data cannot reliably be quantified, reproduced, or reused and ultimately loses scientific value even when it is shared. As such, Microscopy Metadata should be made transparently available to microscope users, automatically collected using community tools [32,71,72,140], and encoded using shared metadata frameworks [28,29].
- **Ensure that instrument maintenance and quality assessment are adequately supported to ensure that they become common practice** at all core facilities and individual laboratories utilizing microscopes and regardless of local resource availability. Specific funding mechanisms should be considered to provide the necessary instrumentation, training and personnel or traveling metrology services for under-resourced areas. This will allow for the performance of instruments to be evaluated at regular intervals using community-defined metrology standards and QC procedures that are appropriate for each experimental question [44–46]. Additional metrics might need to be collected for specific types of experimental approaches and desired outcomes.

- **Emphasize large infrastructure investments in core facilities and regional infrastructural hubs** [97] who employ trained personnel, including bioimaging scientists, data stewards, image analysts, and research software engineers. Such shared infrastructure would increase efficiency and reduce costs by maintaining and assessing the performance of instruments; promote the dissemination of technological advances (hardware and software); facilitate user training; provide guidance for experimental procedures, image analysis, and data stewardship; and provide image analysis and Research Data Management and Sharing (RDMS) services to facilitate the deposition of FAIR data packages containing the appropriate image metadata to specialized bioimage repositories. These infrastructure investments should include providing legal support to review data, identify appropriate Creative Commons (CC) and Open-Source Software (OSS) licensing, and/or carry out Personal Information (PI) redaction (human subject data).
- **Invest in the development of open-hardware devices** (e.g., robotic devices, fluidics systems, environmental control devices, microscopes, etc.) to carry out all aspects of data generation as the most appropriate way to democratize advanced technology and to ensure the efficient use of resources.

**Data Processing and Analysis Metadata Recommendations**

Once acquired, bioimages often require complex processing, visualization, and analysis to extract quantitative information about signal intensity, as well as the location, morphology, association, and movement of biological entities. To ensure that bioimage processing and analysis results are reproducible and FAIR-ready, community-defined guidelines should be followed. In particular:

- Image processing and analysis workflows should be shared as binary containers (e.g., Docker, Singularity, or Podman) that include the complete software environment, ensuring identical pipeline execution for all users.
- The workflow should specify which aspects of microscopy acquisition (e.g., magnification, resolution, signal intensity) may influence processing and should be encoded in the metadata accordingly.
- Every component of the processing and analysis pipeline—including data structure and size, rendering and processing steps, algorithm versions and input parameters, and computing and networking requirements—must be documented in both human- and machine-readable formats to guarantee interpretability, reproducibility, and downstream reuse. This metadata should reside either within the image data file (such as image rendering and processing information) or in the workflow documentation (such as algorithm versions and parameters).
- Therefore, it is essential to capture analysis metadata using community-defined specifications and storage frameworks. This approach maximizes efficiency in extracting and tracking information across all pipeline stages without requiring repeated time-consuming access to the image data itself.

**<u>Data Stewardship Recommendations</u>**

- Responsible stewardship of data and associated metadata throughout the entire lifecycle of quantitative bioimaging experiments is essential to ensure rigor, reproducibility, and the production of high-quality bioimage data that are interpretable and ready for reuse in accordance with FAIR principles.
- Generating and stewarding FAIR image data requires full transparency, management, and reporting of all information related to data generation (including experimental conditions, sample preparation, and image acquisition) as well as processing and analysis (such as image analysis and visualization).
- RDMS cyberinfrastructure that supports the generation and pre-publication stewardship of high-quality FAIR image data should be accessible to all biomedical researchers using microscopes, serving as a crucial prerequisite for bioimage data sharing and reuse.
- This bioimaging RDMS cyberinfrastructure must include advanced computing resources and data repositories that integrate data stewardship, metadata annotation, visualization environments, processing pipelines, and analysis routines (including AI/ML), all connected through high-speed networks to facilitate efficient data transfer.
- Such RDMS cyberinfrastructure is best supported by user-friendly, enterprise-grade, robust, and continuously maintained open-source software that covers all steps of the bioimaging pipeline—from experimental management (e.g., LIMS or ELN), through image acquisition (e.g., Micro-Manager, Pycro-Manager, Python Microscope), to processing, visualization, analysis, and data workflows (e.g., CellProfiler, Fiji, napari, CellPose).
- When proprietary software and instruments are necessary, community-defined standards must be applied to ensure transparency of all relevant algorithms, input parameters, and instrument configurations and performance.
- Metadata collection and reporting must follow community-defined standards at two closely connected levels:
    - **Human-readable**, primarily documented in Materials and Methods [4,20,21,57], representing a subset of Bioimage Metadata that allows users to understand and describe the bioimaging experiment.
    - **Machine-readable**, fully captured in Bioimage Metadata, which includes all information needed to trace the lineage (provenance) and quality of bioimage data, ensuring comprehensive quality, reproducibility, and reusability [28,29,95] .

- **To ensure Machine Readability:**
    - **Metadata should be encoded in community-specified frameworks** to be associated with standardized bioimage data file formats [28,29,95], or workflow documentation, and equipped with readily available software API to facilitate the transfer of information across the different steps

of the bioimaging pipeline. For example, it should be possible to be put in an SQL database automatically. A potential framework that could fulfill these requirements is LinkML[92,141]. It allows for easy authoring of metadata schemas in the YAML format, which can be exported into other formats. It is also able to create classes in various programming languages that can serve to validate metadata.

  - It is imperative to use **specific annotation tools and automation** at all aspects of the bioimaging pipeline to ensure that image processing, visualization, and analysis pipelines can leverage metadata. Particular emphasis should be given to (i) automated processes for microscope systems and peripheral components, including community-defined QC procedures to ensure optimal instrument performance [30,58]; (ii) automated metadata annotation at all phases of the image-data life cycle [32,71,72]; and (iii) integrated image processing, visualization, and analysis pipelines.
  - **Metadata annotations should be backed by ontologies and knowledge graphs.** Ontologies provide descriptions of the hierarchical relationship between concepts and can be used to make metadata machine-readable. Their role in the harmonization of knowledge and data in biomedicine is increasingly recognized, especially in the context of AI, where they can provide valuable constraints that make ML more efficient [104]. Tools such as the Ontology Look Up Service (OLS) [103] can be used to locate appropriate ontological terms, although further development is needed to help sort through duplicates and identify which terms should be used in different contexts.
- Quality, transparency, reproducibility, and reusability require standards defined by all bioimaging community stakeholders.
  - Recent advancements have made it clear that the community, when participating in bioimaging organizations, networks, and initiatives, is willing and ready to take on this challenge.
  - However, deliberate, direct, and targeted funding is needed to ensure that ongoing standardization efforts can be expanded to cover all essential aspects of the bioimaging pipeline.
- In addition to being essential for generating standards, community organizations are ideal ways to promote the broader dissemination and adoption of standards via education, training, and outreach. As such, novel funding models should be implemented to ensure the sustainability of these essential endeavors, promoting the reuse and advancement of quantitative bioimaging and science as a whole.
- Cyber infrastructure for RDMS will impact the economy beyond the biological research enterprise. Tested and trusted microscope QC protocols, and commonly accepted and unambiguous reporting criteria will be beneficial to the many applications of quantitative bioimaging for cellular analysis, including pathology, cell therapies, and regenerative medicine.

## Supplemental Table 1

A non-exhaustive list of relevant communities and initiatives.

| Type | Name | Link |
|---|---|---|
| European FAIR data and service infrastructure | European Open Science Cloud | https://www.eosc.eu/ |
| European imaging data initiative | EUCAIM: EUropean Federation for CAncer IMages | https://www.eibir.org/projects/eucaim/ |
| Federation of Scientific Societies | FASEB Dataworks | https://www.faseb.org/data-management-and-sharing |
| German National Scientific Data Infrastructure | Multi Disciplinary (Data Science, BioImage, etc. etc.) | https://www.nfdi.de/ |
| International community | ABRF: Association of Biomolecular Resource Facilities - Committee on Core Rigor and Reproducibility (CCoRRe) | https://www.abrf.org; https://www.abrf.org/core-rigor-and-reproducibility-ccorre- |
| International community | African BioImaging Consortium (ABIC) | https://www.africanbioimaging.org/ |
| International community | AI4Life: AI models and methods for the life sciences (image data) | https://ai4life.eurobioimaging.eu/ |
| International community | BioImaging North America (BINA)<br>Quality Control and Data Management working group and AIMM interest working group | https://www.bioimagingnorthamerica.org/ |
| International community | Global BioImaging | https://globalbioimaging.org |
| International Community | Human BioMolecular Atlas Program | https://portal.hubmapconsortium.org/ |
| International Community | IBEX Imaging Community | https://ibeximagingcommunity.github.io/ibex_imaging_knowledge_base/ |
| International community | Latin American Bioimaging (LABI) | https://labi.lat/ |
| International community | NEUBIAS - Network of European BioImage Analysts/SoBIAS - Society for Bioimage Analysis | https://eubias.org/NEUBIAS/ |
| International community | Open Microscopy Environment (OME) | https://www.openmicroscopy.org/ |
| International community | QUAREP-LiMi | https://quarep.org/ https://quarep.org/working-groups/wg-7-metadata/ |
| International community | vEM: Volume Electron Microscopy | https://www.volumeem.org/#/ |
| International imaging infrastructure (open access) | Euro-BioImaging ERIC | www.eurobioimaging.eu |
| National/International community | Canada BioImaging | https://www.canadabioimaging.org/ |
| National/International community | I3D:bio - Information Infrastructure for BioImage Data initiative (Germany) | https://www.i3dbio.de |
| National imaging data initiative | NCI Imaging Data Commons (USA) | https://portal.imaging.datacommons.cancer.gov/ |
| National imaging data initiative | NFDI4BIOIMAGE (Germany) | https://nfdi4bioimage.de |
| National image data initiative | RDMS4mic (Germany) | https://german-bioimaging.github.io/RDMS4mic.github.io/ |